\documentclass[12pt]{article}
\usepackage{amsmath}
\usepackage{graphicx,psfrag,epsf}
\usepackage{enumerate}
\usepackage{natbib}
\usepackage[table]{xcolor} 
\usepackage{slashbox}
\usepackage{lineno}

\newcommand{\blind}{1}

\newtheorem{theorem}{Theorem}

\newtheorem{Remark}{Remark}

\begin{document}

\def\spacingset#1{\renewcommand{\baselinestretch}%
{#1}\small\normalsize} \spacingset{1}


\if1\blind
{
  \title{\bf Multiplicative local linear hazard estimation and best one-sided cross-validation}
  \author{Mar\'ia Luz G\'amiz\\
    Department of Statistics and O.R., University of Granada\\
		Mar\'ia Dolores Mart\'inez-Miranda\thanks{
    The authors gratefully acknowledge the support from the Spanish Ministry of Economy and Competitiveness, through grant number MTM2016-76969P, which includes support from the European Regional Development Fund (ERDF), and thank the Centro de Servicios de Inform\'atica y Redes de Comunicaciones (CSIRC), University of Granada, for providing the computing time.}\hspace{.2cm}\\
    Department of Statistics and O.R., University of Granada\\
    Jens Perch Nielsen \\
    Cass Business School, City University of London}
  \maketitle
} \fi

\if0\blind
{
  \bigskip
  \bigskip
  \bigskip
  \begin{center}
    {\LARGE\bf  Multiplicative local linear hazard estimation and best one-sided cross-validation}
\end{center}
  \medskip
} \fi

\bigskip
\begin{abstract}
This paper develops detailed mathematical statistical theory of a new class of cross-validation techniques of local linear kernel hazards and their multiplicative bias corrections. The new class of cross-validation combines principles of local information and recent advances in indirect cross-validation.  A few applications of cross-validating multiplicative kernel hazard estimation do exist in the literature. However, detailed mathematical statistical theory and small sample performance are introduced via this paper and further upgraded to our new class of best one-sided cross-validation. Best one-sided cross-validation turns out to have excellent performance in its practical illustrations, in its small sample performance and in its mathematical statistical theoretical performance.
\end{abstract}

\noindent
{\it Keywords:} {Aalen's multiplicative model; Multiplicative bias correction; 
Bandwidth; Indirect cross-validation.

\vfill

\newpage
\spacingset{1.45} 

\section{Introduction}
\label{sec:intro}
\label{intro}

There is a growing interest in validation techniques.  While validation was always a crucial element of mathematical statistics, the use of validation techniques are growing rapidly at the moment under labels such as big data, machine learning or artificial intelligence. Many of these developments seem less patient with laborious mathematical statistical model formulation and estimation theory than what has been the trademark of the field of mathematical statistics.  Instead inspiration seem to be taken from neighbouring fields  such as engineering, computer science, public health or actuarial science where specific knowledge is present on the problem at hand allowing the development of clever and perhaps computationally challenging algorithms often replacing more labour intensive procedures of the past. These algorithms are often defined in such a way that they can change and learn over time via some optimization criteria and an efficient validation procedure. One example of such work relevant to the work of this paper is the paper of \cite{Munoz:Laan:12}, where an impressive algorithm is developed to solve a complicated survival problem. The introduced methodology is inspired by machine learning calling its validation procedure for a Super Learner. However, while the Super Learner is optimal in some sense, see \cite{Laan:etal:07},  then it is not optimal in the more detailed mathematical statistical sense that we consider in this paper. And this is not only because \cite{Munoz:Laan:12} consider piecewise constant hazard models that are less efficient than kernel smoothers. It is also because the validation theory presented in the paper does not provide the mathematical detail promoted in this paper and therefore crucial insight of noisy second order components is not included in the theory. The approach of \cite{Munoz:Laan:12} is just one among many machine learning inspired survival analyses approaches. This paper will consider one dimension only. Multidimensional cross-validation and one-dimensional cross-validation are closely related and mathematical definitions are similar. However, even in the one-dimensional case we face challenging theoretical as well as practical issues with cross-validation being too noisy and unstable and to such an extend that we cannot any longer recommend cross-validation in one dimension without some amendment for the noise involved. Our intention is that multidimensional big data type of problems, with further issues with data sparsity and noisy cross-validation, should benefit in the future from the insight on cross-validation analyses as provided in this paper. The mathematical point of view of this paper was perhaps initiated via the early contribution of \cite{Hall:Marron:87} that provided a decision theoretical framework to distinguish between plug-in estimators, aiming at minimizing a mean integrated square error, and cross-validation aiming at minimizing the infeasible stochastic integrated square error. They concluded that plug-in did better from an asymptotic perspective even when the aim was the explicit aim of cross-validation: to get as close as possible to the infeasible minimization of the integrated squared error. One could view this as the foundation of a new decision theoretical framework to understand the quality of kernel bandwidth selection; a tractable place to start when understanding the complicated world of model selection.  \cite{Hall:Johnstone:92} pointed out that for any bandwidth selector there are two sources of noise for kernel density estimation, one that one can never get rid of and another one that seems to differ for different methods. The second source of noise could theoretically go as low as to zero such that one was left with the first noise component as a lower bound on noise.  The plug-in type of methods had considerable lower second-component-noise than cross-validation and plug-in was very popular in practice in the nineties with \cite{Sheather:Jones:91} being the perhaps most popular single method. However, plug-in methods depend on complicated underlying mathematical detail and it does not easily generalise to new problems in the same straightforward way as cross-validation does. This is perhaps the single most important reason why cross-validation has regained its importance and is used for a wide variety of complicated problems in mathematical statistics, big data, machine learning and artificial intelligence. \cite{Hart:Yi:98} introduced the concept of indirect cross-validation - formulated in nonparametric kernel regression - that simply meant that cross-validation was performed on an alternative kernel and the bandwidth was scaled back to the original kernel used for estimation. \cite{Hart:Yi:98} suggested to use one-sided kernels as the alternative kernels because of their good practical performance and simple rescaling.  In density estimation \cite{Savchuk:etal:10} suggested a clever combination of a normal-bandwidth kernel and an oversmoothed kernel as alternative kernel to achieve the same mathematical statistical asymptotic performance as the plug-in estimator without the need of a pilot. However, there was one catch with the elegant approach of \cite{Savchuk:etal:10}. Their approach needed to estimate some tuning parameters to decide the relative weight of the oversmoothed kernel that was contributing to the asymptotic noise via some term of lower order. So, even though \cite{Savchuk:etal:10} in principle did pilot free estimation then there was still some tuning going on and some extra terms of just slightly lower order. And that was perhaps exactly the problem of the original plug-in methods  as in  \cite{Sheather:Jones:91}: that something with lower order noise had to be estimated - the pilot - and terms of slightly lower order had to be ignored in the asymptotic results.  In this paper we define three dogmas for a cross-validation estimator:  
\begin{enumerate}
\item It should be a direct estimation based on principles without complicated mathematical adjustments.
\item Extra terms of slightly lower order are not allowed in the expansions.
\item Further smoothing than those necessary for the original estimator is not allowed to be assumed while analysing the quality of the bandwidth selector.
\end{enumerate}

The original cross-validation estimator and the approach of \cite{Hart:Yi:98} lives up to all three dogma rules while the plug-in type estimators of e.g. \cite{Sheather:Jones:91} and \cite{Savchuk:etal:10} violate all three.  We believe this to be the reason why \cite{Savchuk:etal:10}  could not make their new pilot-free bandwidth selector convince in practice, and why \cite{Mammen:etal:11}, \cite{Mammen:etal:14} concluded that their double one-sided kernel density bandwidth selector - directly inspired by \cite{Hart:Yi:98} - worked better in practice than the  estimators of \cite{Sheather:Jones:91} and \cite{Savchuk:etal:10}. The fundamental principles of this paper is therefore the three dogmas above and the decision theoretical framework of \cite{Hall:Marron:87} and this has let us to explore double one-sided cross-validation and one-sided cross-validation even further because of their apparent practical superiority on the market of current kernel bandwidth selectors. A detailed investigation of both sides of local one-sided bandwidth selection showed us a perhaps surprising fact. While the left-side and the right-side cross-validation procedures have the same mathematical statistical behaviour, they do perform very differently in practice. Often one of the two sides breaks down completely. Therefore one-sided cross-validation does not really work in practice, it breaks down too often. Double one-sided cross-validation works better than one-sided cross-validation in a wide variety of kernel smoothing problems, see for example Mammen et al. (2011,2014), G\'amiz et al. (2013a,b, 2016). A closer investigation going through local features of individual simulation samples reveals that behind a good double-one-sided cross-validation result often hides an average of a good one-sided estimator and a somehow random result from the other side. Partly because of some prior knowledge keeping the bandwidth search in a reasonable interval and partly because of pure luck, the simulation results of double one-sided cross-validation are often very good indeed. The suggestion of this paper is to improve the stability of one-sided cross-validation via a local information principle inspecting at every single local point whether to use the right-side or the left-side for cross-validation. This approach is indeed very stable in its practical performance, it obeys the three above dogmas and it provides the exact same asymptotic performance as its less stable one-sided and double-sided competitors mentioned above. We call the new approach best one-sided cross-validation. This paper furthermore introduces the mathematical statistical approach of \cite{Hall:Marron:87} to multiplicatively bias corrected local linear kernel hazard estimators and it introduces asymptotic theory and practical implementation of  best-one-sided-cross-validation for these multiplicatively bias corrected hazard estimators.  Multiplicative bias correction is known to improve the practical implementation of kernel hazard estimation, see \cite{Nielsen:98} and \cite{Nielsen:Tanggaard:01}. This parallels insights from the more researched world of kernel density estimation, see for example Jones, Linton and Nielsen (1995) and \cite{Jones:Signorini:97}. The latter went through a series of small sample studies of kernel density estimation procedures to conclude that multiplicative bias correction seemed to be the best. The contribution of this paper is therefore also to update mathematical statistical theory and practice to the perhaps best practically performing  kernel hazard estimator we have: the local linear multiplicatively bias corrected kernel hazard estimator.

The rest of the  paper is organized as follows. In Section \ref{sec:training and learning} we describe the link between our proposal and methods in machine learning. In Section  Section \ref{sec:model} we formulate the model we assume in the paper and present two hazard estimators namely the local linear estimator and its multiplicative bias correction. Bandwidth selection for these estimators through cross-validation and the double one-sided cross-validation of  \cite{Gamiz:etal:16} method is described in Section  \ref{sec:bo}, and our new best one-sided cross-validation method is suggested. The asymptotic properties of all presented validated bandwidths are analysed in Section \ref{sec:asym}. Assumptions and proofs are provided in the supplementary material. Two case studies show the applicability of our proposals, which are described in Section \ref{sec:data}. In Section \ref{sec:sim} we describe simulation experiments to evaluate the finite sample properties of our proposal. All numerical calculations have been performed with \texttt{R}. Best one-sided cross-validation is implemented in the \texttt{DOvalidation} package (\cite{Gamiz:etal:17}), along with double one-sided cross-validation and cross-validation, for the local linear hazard estimator and its multiplicative bias correction.

\section{Training and learning versus cross-validation and adjusted cross-validation}
\label{sec:training and learning}

To motivate our research beyond a wider crowd than experts in nonparametric hazard estimation, our point of view is formulated below via standard vocabulary from machine learning and artificial intelligence. Let us assume we observe $n$ individuals over some time that could potentially be filtered via truncation and censoring and  let $A$ be a training set and $B$ be a learning set such that the two sets united equals the set $\{1,...,n\}$. Let for the purpose of a discussion the number of elements of $A$ be 80\% of $n$ and the number of  elements in $B$ be 20\% of $n$. Then a standard approach to validation, see again \cite{Munoz:Laan:12}, would be to estimate the hazard on the training set and evaluate it via the learning set. Under some standard independence assumptions this will lead to a decrease in efficiency of estimation itself corresponding to ignoring 20\% of the data set and it will decrease the efficiency on the validation approach -- compared to cross-validation and the theoretical approach considered in this paper -- corresponding to ignoring 80\% of the data set. One could of course consider all possible combinations of training and learning sets and average all these validations into one single validation principle or learning principle. This would correspond to a computationally inefficient cross-validation. In conclusion: even if all possible combinations of trainers and learners are calculated, we end up with standard cross-validation with the well known problems of data sparsity and noise. With the help of the theory originally developed by \cite{Hall:Marron:87} in the kernel density context, we will in this paper -- in the kernel hazard context -- consider more efficient use of data when estimators are validated or when trainers are learning. It turns out that this is indeed possible via relatively straightforward adjustments of standard cross-validation.

\section{The counting process model and kernel hazard estimators}
\label{sec:model}

In this section we formulate events via counting processes. Counting processes are well designed when  event data are filtered for example via truncation or censoring. An individual zero-one valued exposure process simply keeps tracks on whether an individual is under risk or not at any particular point in time. We assume that individuals are independent and that data filtering is non-informative. Formally, we observe $n$ individuals, $i = 1,\ldots,n$. Let $N_i$ count observed failures for the $i$th individual in the time interval $[0,T]$, $N_i$ can take values $0$ or $1$. We assume that $N_i$ is a one-dimensional counting process with respect to  an increasing, right continuous, complete filtration ${\cal F}_{t}$, $t \in [0, T]$, i.e., it obeys \emph{less conditions habituelles}, see Andersen et al. (1993, pp. 60). We assume Aalen's multiplicative model (\cite{Aalen:78}) where the random intensity is written as, 
$
\lambda_i(t)=\alpha(t) Y_i(t),
$
with no restriction on the functional form of the hazard function $\alpha(\cdot)$. Here $Y_i$ is a predictable process taking values $0$ or $1$, indicating (by the value 1) when the $i$th individual is at risk and under observation. We assume that $\left(N_1,Y_1\right),\ldots,\left(N_n,Y_n\right)$ are i.i.d. for the $n$ individuals. With these definitions $\lambda_i$ is predictable and the processes $M_i(t)=N_i(t) - \Lambda_i(t)$, $i=1,\ldots,n$, with $\Lambda_i(t)=\int_0^t \lambda_i(s) \ ds$, are squared integrable local martingales. 

As an example we illustrate how the above stochastic processes look like in the case of independent and non-informative left truncation and right censoring, where $n$ tuples  $(L_i,Z_i,\delta_i)$, $i=1,..,n$,  are observed.  Here $L_i$ is the time the $i$th individual enters the study; $Z_i$ is the time $i$th individual leaves the study either because an event has happened or because of right censoring; and $\delta_i$ is binary and equal to 1 if an event -- for example death or an onset of a disease --  is the reason for the $i$'th individual to leave the study and the value is zero when the reason for the $i$th individual to leave the study was uninformative right censoring.  In this case, the process $Y_i$ above would be $Y_i(t) = I(L_i \leq t < Z_i)$ and $N_i(t)= I(Z_i \leq t)\delta_i$, where $I(\cdot)$ is the indicator function. Hereafter we will work in the convenient and general stochastic process formulation only.

The local linear kernel hazard estimator in our general stochastic process formulation was introduced by \cite{Nielsen:Tanggaard:01} and it is defined as 
\begin{equation}\label{LL}
\widehat{\alpha}_{b,K}^{{\rm LL}}(t)=\sum_{i=1}^n \int_0^T \bar{K}_{t,b}(t-s)dN_i(s),
\end{equation}
with the stochastic local linear kernel
\begin{equation}\label{Kbar}
\bar{K}_{t,b}(t-s)=\frac{a_{2,K}(t)-a_{1,K}(t)(t-s)}{a_{0,K}(t)a_{2,K}(t)-\{a_{1,K}(t)\}^2}  K_b\left ( {t-s} \right ),
\end{equation}
where  $K_b(u)= b^{-1} K(u/b)$ and
$a_{j,K}(t)=\int_0^T K_b\left ( t-s \right )(t-s)^j Y(s)ds$, for $j=0,1,2$.  Here $K$ is a kernel function with support $[-1,1]$ and $b>0$ is the bandwidth parameter.

The local linear kernel $\bar{K}_{t,b}$ satisfies the properties: $\int_0^T \bar{K}_{t,b}(t-s)Y(s)ds=1$, $\int_0^T \bar{K}_{t,b}(t-s)(t-s)Y(s)ds=0$ and $\int_0^T \bar{K}_{t,b}(t-s)(t-s)^2Y(s)d s>0$. Thus, $\bar{K}_{t,b}$ can be interpreted as a second order kernel with respect to the stochastic measure $\mu$, where $d\mu(s)=Y(s)ds$, and $Y(t) =\sum_{i=1}^n Y_i(t)$ is the aggregated risk process. Defining the aggregated failure process, $N(t)=\sum_{i=1}^n N_i(t)$, we can write $\widehat{\alpha}_{b,K}^{{\rm LL}}(t)=\int_0^T \bar{K}_{t,b}(t-s)dN(s)$.

The multiplicative bias corrected (MBC) estimator constructed from the local linear hazard estimator is defined as
\begin{equation}\label{MBC}
 \widehat{\alpha}_{b,K}^{{\rm MBC}}(t)=\sum_{i=1}^n \int \bar{K}^{{\rm MBC}}_{t,b}(t-s) \widehat{\alpha}_{b,K}^{{\rm LL}}(t)\{\widehat{\alpha}_{b,K}^{{\rm LL}}(s)\}^{-1}dN_i(s) ,
\end{equation}
where the MBC kernel is 
\begin{equation}\label{K.M}
\bar{K}^{{\rm MBC}}_{t,b}(t-s)=\frac{a^{\rm MBC}_{2,K}(t)-a^{\rm MBC}_{1,K}(t)(t-s)}{a^{\rm MBC}_{0,K}(t)a^{\rm MBC}_{2,K}(t)-\{a^{\rm MBC}_{1,K}(t)\}^2} \left\{\widehat{\alpha}_{b,K}^{{\rm LL}}(s)\right \}^2 K_b\left ( {t-s} \right ) ,
\end{equation}
with $a^{\rm MBC}_{j,K}(t)=\int_0^T K_b\left ( t-s \right )(t-s)^j \left\{\widehat{\alpha}_{b,K}^{{\rm LL}}(s)\right\}^2 Y(s)ds$, for $j=0,1,2$.

\section{Cross-validation and best one-sided cross-validation of our two estimators}
\label{sec:bo}

The two kernel hazards estimators  considered in this paper depend on a bandwidth parameter that determines the smoothness degree of the resulting estimates. Choosing the bandwidth parameter is a crucial problem that starts by defining what the optimal bandwidth would be, so it can be estimated from data. 

Let $\widehat{\alpha}_{b,K}$ denote a kernel hazard estimator with bandwidth $b$ and kernel $K$, which can be any of the two defined in (\ref{LL}) or (\ref{MBC}). Ideally we would like a bandwidth parameter $b$ that minimizes the integrated squared error (ISE) given by
\begin{equation*} 
\Delta_K(b)=n^{-1}\sum_{i=1}^n \int_0^T \left\{\widehat{\alpha}_{b,K}(s)-\alpha(s)\right\}^2Y_i(s)w(s)ds,
\end{equation*}
where $w(\cdot)$ is some weight function. However, the minimizer of the ISE,  $\widehat b_{{\rm ISE},K}$, depends on the unknown hazard function and it is infeasible in practice. In this paper we consider $\widehat b_{{\rm ISE},K}$ as the optimal bandwidth and in this section we present estimates based on the cross-validation method. We refer the reader to \cite{Gamiz:etal:16} for the history of cross-validation in kernel hazard estimation based on counting processes.

First notice that minimizing  $\Delta_K(b)$  is equivalent to minimizing
$$
n^{-1}\left[\sum_{i=1}^n \int_0 ^T\left  \{\widehat{\alpha}_{b,K}(s)\right \}^2Y_i(s)w(s)ds- 2\sum_{i=1}^n \int_0^T \widehat{\alpha}_{b,K}(s)\alpha(s)Y_i(s)w(s)ds \right],
$$
and only the second term depends on the unknown hazard. The cross-validation approach estimates this second term from the data replacing ${\alpha}(s)ds$ by its empirical counterpart ${dN_i(s)}$. The cross-validated bandwidth, denoted by 
$\widehat b_{{\rm CV},K}$, is therefore the minimizer of 
\begin{equation} \label{cv_score}
\widehat{Q}_K(b)=n^{-1}\left[\sum_{i=1}^n \int_0 ^T \left\{\widehat{\alpha}_{b,K}(s)\right\}^2 Y_i(s)w(s)ds 
                              - 2\sum_{i=1}^n \int_0^T \widehat{\alpha}_{b,K}^{[i]}(s)w(s)dN_i(s) \right ],
\end{equation}
where $\widehat{\alpha}_{b,K}^{[i]}(s)$ is the estimator arising when the data set is changed by setting the stochastic process $N_i(s)$ equal to 0 for all $s \in [0,T]$.

A practical and theoretical improvement of cross-validation was given in \cite{Gamiz:etal:16} that developed double one-sided cross-validation (DO-validation), as a simple average of two indirect cross-validated bandwidths. Indirect cross-validation makes use of the fact that, under mild regularity conditions, asymptotically optimal bandwidths for two kernel estimators with different kernels $K$ and $L$ differ by a factor that only depends on the two kernels $K$ and $L$. In indirect cross-validation one applies cross-validation to a kernel estimator with kernel $L$ and afterwards one multiplies the cross-validation bandwidth by the factor (depending on $K$ and $L$) to get a bandwidth for the kernel estimator with kernel $K$. Such a construction makes sense if cross-validation for a kernel estimator with kernel $L$ works better than cross-validation for a kernel estimator with kernel $K$. Double one-sided cross-validation averages the two indirect cross-validation bandwidths  based on one-sided kernels: the left-sided $K_{{\rm L}}(u)=2K(u)I(u<0)$, or the right-sided $K_{{\rm R}}(u)=2 K(u)I(u>0)$. More specifically, two one-sided cross-validation criteria, $\widehat{Q}_{K_{{\rm L}}}(b)$ and $\widehat{Q}_{K_{{\rm R}}}(b)$,  are defined as in \eqref{cv_score} but replacing $K$ with $K_{{\rm L}}$ and $K_{{\rm R}}$, respectively. Denoting by $\widehat{b}_{{\rm CV},K_{\rm L}}$ and $\widehat{b}_{{\rm CV},K_{\rm R}}$ their minimizers, the  double one-sided cross-validation bandwidth estimate is the (conveniently) weighted average of these:
$$
\widehat b_{{\rm DO},K} = \frac{1}{2} \rho \left (\widehat{b}_{{\rm CV},K_{{\rm L}} }+ \widehat{b}_{{\rm CV},K_{{\rm R}}} \right).
$$
 For the local linear hazard estimator defined in ($\ref{LL}$), the factor $\rho$ is given by
\begin{equation}\label{roLL}
\rho^{\rm{LL}}=\left ( \displaystyle\frac{R(K)}{R(\bar{K}^*_{{\rm L}} )} \displaystyle%
\frac{\mu_2(\bar{K}^*_{{\rm L}} )^2}{\mu_2(K)^2} \right )^{1/5}.
\end{equation}
Here, for a general kernel $L$,  $\bar L^*$ denotes the equivalent local linear kernel defined as
\begin{equation}\label{KLequiv}
\bar{L}^*(u)=\frac{\mu_2\left(L\right)-\mu_1\left(L\right)u}{\mu_2\left(L\right)-\mu_1\left(L\right)^2}L(u),
\end{equation}
where $\mu_2(L)=\int u^2L(u)du$ and $R(L)= \int L^2(u)du$. Notice that $\bar L^*=L$ if $L$ is symmetric.

For the MBC estimator, $\widehat{\alpha}_b^{\rm MBC}$, defined in ($\ref{MBC}$), the factor $\rho$ becomes
\begin{equation}\label{roM}
\rho^{\rm MBC}=\left ( \displaystyle\frac{R(\Gamma_K)}{R(\Gamma_{\bar{K}^*_{{\rm L}}})} \displaystyle%
\frac{\mu_2(\bar{K}^*_{{\rm L}})^4}{\mu_2(K)^4} \right )^{1/9}.
\end{equation}

 The asymptotic theory developed in \cite{Gamiz:etal:16} for the local linear hazard estimator  showed that left- and right-sided cross-validation have the same asymptotic properties, but different finite sample  performance.  There are situations where one of the two one-sided cross-validation methods breaks down so the averaging strategy of double one-sided cross-validation becomes inappropriate. The natural reaction in these cases would be to take the side which is working fine. One common reason for one of the two one-sided cross-validated bandwidths to break down is the lack of occurrences (or exposures) in one of the two directions. Perhaps because of a boundary.  Best one-sided cross-validation (BO-validation)  introduced in this paper simply uses the one-sided version that, via local information, is predicted to work best at every single point $t$. There can therefore be both left-sided and right-sided kernels involved in best one-sided cross-validation. Imagine for example that the estimation interval is $(0,1)$ such that two boundaries are present then one would expect to use different sided kernels for a $t$ close to the left boundary $0$ and for a $t$ close to the right boundary $1$. 

 For the local linear hazard estimator we define the kernel estimator needed for best one-sided cross-validation as
\begin{equation}\label{BO_LL}
\widehat{\alpha}^{{\rm BO, LL}}_{b,K}(t)=\int_0^T \left\{\bar{K}_{t,b;{\rm L}}(t-s) \xi_b(t)+ \bar{K}_{t,b;{\rm R}}(t-s)\left(1-\xi_b(t)\right)\right\}dN(s)
\end{equation}
where $\bar{K}_{t,b;{\rm L}}$ and $\bar{K}_{t,b;{\rm R}}$ are respectively the left and right versions of the local linear kernel $\bar{K}_{t,b}$ in (\ref{Kbar}), and $\xi_b(t)$ is a stochastic function, depending on the estimation time $t$ and the bandwidth $b$, which takes the value 1 when the ``best'' side to consider is the indicated by the kernel $K_{\rm L}$, and the value 0 otherwise. The combination of one-sided kernels that appears in the integrand of expression ($\ref{BO_LL}$) is a kernel function which we denote as
\begin{equation}\label{BOker}
\bar{K}^{{\rm BO, LL}}_{b,K}(t-s)=\bar{K}_{t,b;{\rm L}}(t-s) \xi_b(t)+ \bar{K}_{t,b;{\rm R}}(t-s)\left(1-\xi_b(t)\right).
\end{equation}
Thus we write the estimator as $\widehat{\alpha}^{{\rm BO, LL}}_{b,K}(t)=\int_0^T \bar{K}^{{\rm BO, LL}}_{b,K}(t-s)dN(s)$.
  
For each time $t$, to designate which side is ``best'', $\xi_b(t)$ can be defined in terms of the occurrence process by
\[
\xi_b^{\rm O}(t)=I\left (\int_{t-b}^tdN(s) > \int_{t}^{t+b} dN(s)\right),
\]
or  the exposure process by
\begin{equation} \label{xi.E}
\xi_b^{\rm E}(t)=I\left( \int_{t-b}^t Y(s) ds > \int_{t}^{t+b} Y(s)ds\right).
\end{equation}

With any of these $\xi_b^{\rm O}$ or $\xi_b^{\rm E}$, the best one-sided cross-validation bandwidth estimate is defined as
\begin{equation} \label{bo}
\widehat{b}^{LL}_{{\rm BO},K}=\rho^{\rm LL} \arg\min_b \widehat Q_K^{{\rm BO, LL}}(b),
\end{equation}
 where $\widehat Q_K^{{\rm BO, LL}} $ is the cross-validation score in (\ref{cv_score}) calculated with the kernel estimator $\widehat{\alpha}^{{\rm BO, LL}}_{b,K}(t)$, defined in (\ref{BO_LL}). In a similar way we define the best one-sided cross-validation bandwidth estimate for the MBC hazard estimator, $\widehat{b}^{MBC}_{{\rm BO},K}$,  as in (\ref{bo}) but replacing the factor $\rho^{\rm LL}$ with  $\rho^{\rm MBC}$, given in (\ref{roM}), and defining the best one-sided cross-validation score, $\widehat Q_K^{{\rm BO, MBC}} $, with the hazard estimator
\begin{equation}\label{BO_MBC}
\widehat{\alpha}^{{\rm BO, MBC}}_{b,K}(t)=\int_0^T \left\{\bar{K}^{{\rm MBC}}_{t,b;{\rm L}}(t-s)\frac{\widehat{\alpha}^{\rm LL}_{b,K_{\rm L}}(t)}{\widehat{\alpha}^{\rm LL}_{b,K_{\rm L}}(s)} \xi_b(t)+ \bar{K}^{{\rm MBC}}_{t,b;{\rm R}}(t-s)\frac{\widehat{\alpha}^{\rm LL}_{b,K_{\rm R}}(t)}{\widehat{\alpha}^{\rm LL}_{b,K_{\rm R}}(s)}\left(1-\xi_b(t)\right)\right\}dN(s).
\end{equation}

\section{Asymptotic theory}
\label{sec:asym}

In this section we develop theory for the asymptotic behaviour of bandwidth selectors for the local linear hazard estimator and its multiplicatively bias correction. For each estimator we prove the asymptotic normality for bandwidths selectors based on indirect cross-validation, the double one-sided cross-validation estimate by \cite{Gamiz:etal:16} and the new best one-sided cross-validation. Our theoretical results thus extend the results given in \cite{Gamiz:etal:16}, by including the new best one-sided cross-validation for  local-linear hazard estimator and considering the MBC estimator.  

Recall that the ISE of a kernel hazard estimator, $\widehat{\alpha}_{b,L}$, with bandwidth $b$ and general kernel $L$,  was defined as above as
\begin{equation} \label{eq:Q_0}
\Delta_{L}(b)=n^{-1}\int_0^T\left(\widehat{\alpha}_{b,L}(t)-\alpha(t)\right)^2w(t)Y(t)dt,
\end{equation} 
and its minimizer denoted as $\widehat b_{{\rm ISE},L}$. 
Using a convenient expansion of the  ISE and martingale theory we derive  the asymptotic normality of the bandwidth estimates derived for a kernel hazard estimator, $\widehat{\alpha}_{b,K}$, such as the local linear in (\ref{LL}) and the MBC estimator in  (\ref{MBC}). Hereafter we will make explicit reference  to the considered  hazard estimator on the bandwidth estimate, using superscripts (LL or MBC).  Besides a kernel denoted by $K$ is  assumed to be symmetric (see Assumption A1 in supplementary material), while we use the notation $L$ for a general kernel that can be asymmetric, as the one-sided kernels involved in double one-sided cross-validation and best one-sided cross-validation. 

\subsection{A general theorem for indirect cross-validation with a local linear estimator}

%
Let consider the local linear hazard estimator,  $\widehat{\alpha}_{b,L}^{\rm LL}$, given in (\ref{LL}), with bandwidth $b$ and kernel $L$. 
Following the same arguments described in \cite{Nielsen:Tanggaard:01}, the error $\widehat{\alpha}_{b,L}^{\rm LL}(t)-\alpha(t)$, can be decomposed as $\widehat{\alpha}_{b,L}^{\rm LL}(t)-\alpha(t)=V_{b,L}^{\rm LL}(t)+B_{b,L}^{\rm LL}(t)$, where $B_{b,L}^{\rm LL}$ is a stable part converging in probability to zero, given by
\begin{equation}\label{B.LL}
B_{b,L}^{\rm LL}=\int_0^T \bar{L}_{t,b}(t-s) \left(\alpha(s)-\alpha(t)\right)Y(s)ds;
\end{equation}
and $V_{b,L}^{\rm LL}$ is a variable part converging to a Normal distribution, given by
\begin{equation}\label{V.LL}
V_{b,L}^{\rm LL}(t)= \int_0^T \bar{L}_{t,b}(t-s)dM(s).
\end{equation}
Using the above decomposition we can expand the ISE  for the local linear estimator, using standard martingale theory along with the approach of \cite{Mammen:Nielsen:07}. In Lemma 4 in the supplementary material we show that, under some regularity assumptions,  $\Delta_{L}^{\rm LL}(b)$ in (\ref{eq:Q_0}) is asymptotically equivalent to 
\[
M_{L}^{\rm LL}(b)=b^{4}\frac{\mu_2^2\left(\bar{L}^{*}\right)}{4} \int \left\{\alpha''(t)\right\}^2\gamma(t)w(t)dt+(nb)^{-1}R\left({\bar{L}^*}\right)\int \alpha(t)w(t)dt,
\]
where  $\gamma(t) = n ^{-1} E \left[Y(t)\right]$ is the expected exposure function. From this approximation a deterministic optimal bandwidth for the local linear estimator with kernel $L$ is defined as
\begin{equation}\label{bMISE.LL}
b^{\rm LL}_{{\rm MISE},L}=C_{0,L}^{\rm LL} n^{-1/5}  \ \ \ \ \ \mbox{with} \ \ \ \ \ C_{0,L}^{\rm LL}=\left[\frac{R\left({\bar{L}^*}\right) \int \alpha(t)w(t)dt}{ \mu_2^2\left(\bar {L}^*\right) \int \left(\alpha''(t)\right)^2\gamma(t)w(t)dt}\right]^{1/5} .
\end{equation} 
Our main result in this section provides the asymptotic normality of the three bandwidth estimates for the local linear hazard estimator, $\widehat{b}_{{\rm CV},K}^{\rm LL}$, $\widehat{b}_{{\rm DO},K}^{\rm LL}$, and $\widehat{b}_{{\rm BO},K}^{\rm LL}$, as well as the optimal infeasible bandwidth $\widehat{b}_{{\rm MISE},K}^{\rm LL}$. Note that the later is the optimal bandwidth aimed by plug-in bandwidth selection rules. The result is stated in the following theorem and the proof provided in the supplementary material. 

\begin{theorem}\label{Th:local}
Under assumptions A1--A3, the bandwidth selectors, $\widehat{b}_{{\rm BO},K}^{\rm LL}$, $\widehat{b}_{{\rm DO},K}^{\rm LL}$,$\widehat{b}_{{\rm CV},K}^{\rm LL}$, and $\widehat{b}_{{\rm MISE},K}^{\rm LL}$, for the local linear estimator with kernel $K$ satisfy

\begin{eqnarray*}
n^{3/10}\left(\widehat{b}_{{\rm BO},K}^{\rm LL}-\widehat{b}_{{\rm ISE},K}^{\rm LL}\right) &&\longrightarrow N\left(0,S_{2}^{\rm LL}+S_{1}^{\rm LL} \Psi_{{\rm BO},K}^{\rm LL}\right) \\
n^{3/10}\left(\widehat{b}_{{\rm DO},K}^{\rm LL}-\widehat{b}_{{\rm ISE},K}^{\rm LL}\right) &&\longrightarrow N\left(0,S_{2}^{\rm LL}+S_{1}^{\rm LL} \Psi_{{\rm DO},K}^{\rm LL}\right) \\
n^{3/10}\left(\widehat{b}_{{\rm CV},K}^{\rm LL}-\widehat{b}_{{\rm ISE},K}^{\rm LL}\right) &&\longrightarrow N\left(0,S_{2}^{\rm LL}+S_{1}^{\rm LL} \Psi_{{\rm CV},K}^{\rm LL}\right) \\
n^{3/10}\left(\widehat{b}_{{\rm MISE},K}^{\rm LL}-\widehat{b}_{{\rm ISE},K}^{\rm LL}\right) &&\longrightarrow N\left(0,S_{2}^{\rm LL}+S_{1}^{\rm LL} \Psi_{{\rm MISE},K}^{\rm LL}\right)
\end{eqnarray*}
where
\begin{eqnarray*}
S_{1}^{\rm LL}&=& \frac{1}{25}\frac{R\left(K\right)^{-7/5}   \left(\int \alpha^2 (t) w^2(t) \ dt \right)}
{\mu_2(K)^{6/5}\left(\int \alpha''(t)^2 \gamma(t) w(t) \ dt \right) ^{\frac{3}{5}}  \left(\int \alpha (t) w(t) \ dt \right) ^{-7/5} } , \\
&& \\
&& \\
S_{2}^{\rm LL}&=&\frac{4}{25}
\frac{ R\left(K\right)^{-2/5} \left(\int \alpha'' (t)^2 \gamma(t) w^2(t) \alpha(t) \ dt \right)}
{ \mu_2(K)^{6/5} \left(\int \alpha (t) w(t) \ dt \right) ^{2/5} \left(\int \alpha''(t)^2 \gamma(t) w(t) \ dt \right) ^{8/5}},
\end{eqnarray*}
and
\begin{eqnarray*}
 \Psi_{{\rm BO},K}^{\rm LL}&=&\Psi_{{\rm DO},K}^{\rm LL}=  \int{\left \{\frac{R\left(K\right)}{R\left({\bar{ L}^*}\right)}\left(H_{\bar{K}_{\rm L}} - G_{\bar{K}_{\rm L}}\right)\left(\rho^{\rm LL} u\right)-H_{K}(u)\right\}^2}   \ du ,\\
\Psi_{{\rm CV},K}^{\rm LL}&=&  \int \left\{G_{K}( u)\right\}^2  \ du , \\
\Psi_{{\rm MISE},K}^{\rm LL}&=&  \int \left\{H_{K}( u)\right\}^2  \ du , 
\end{eqnarray*}
defining  the functions $G_L(\cdot)$ and $H_L(\cdot)$ as
\begin{eqnarray*}
G_L(w)&=&I\left(w \neq 0 \right) 2\bar L^{\ast}_1(w) ,\\
H_L(w)&=&I\left(w\neq 0\right)\int \bar L^{\ast}(u)\left\{\bar L^{\ast}_1(u+w)+\bar L^{\ast}_1(u-w)\right\}du,
\end{eqnarray*}
 with $\bar L^{\ast}_1(u)=-\bar L^{\ast}(u)-u\bar L^{\ast '}(u)$.
\end{theorem}

\begin{Remark}
\cite{Gamiz:etal:16} pointed out that all bandwidth estimates have similar asymptotics with the only difference of the factor $\Psi_{\cdot,K}^{\rm LL}$. These authors considered three common choices of the kernel $K$ (Epanechninov, quartic and sextic kernels) and calculated the numerical value of this  factor. It allows the  comparison of  the asymptotic performance of bandwidth selectors. These numerical values are reported in the first rows of Table \ref{tab:factors}. Note that these values were multiplied by 2 for convenience in the former paper.

\end{Remark}

\begin{table}[h]
\caption{\label{tab:factors}Comparison of asymptotic variances among bandwidth selection methods. Factors $\Psi_{\bullet,K}^{\bullet}$ defined in Theorems \ref{Th:local}  and \ref{Th:multi}, are shown for the local linear and the MBC estimators, and three common symmetric kernels  $K$: Epanechninov, quartic and sextic. }
	\centering
{
\begin{tabular}{lcccccc} 
		\\\hline\noalign{\smallskip}
        & \multicolumn{3}{c} {Local linear estimator} &\multicolumn{3}{c}{MBC estimator}\\ 
Method  	&  { Epanechnikov}  & {  Quartic } &  {Sextic}	&  { Epanechnikov}  & {  Quartic } &  {Sextic}\\ \hline
BO-validation 	&1.09&0.95&1.18      & 4.41 &  2.44   & 2.05  \\
DO-validation 	&1.09&0.95&1.18      & 4.41 &  2.44   & 2.05  \\
Cross-validation	 & 3.6 & 2.86 & 3.49  &	9.87 &   6.10  & 6.50\\
Plug-in    &  0.36 & 0.46 & 0.59          	& 0.84  &  0.95  & 1.31       \\\hline
\end{tabular}
}
\end{table}


\subsection{A general theorem for indirect cross-validation with a multiplicatively bias corrected estimator}


Consider now the MBC estimator defined in (\ref{MBC}),   $\widehat{\alpha}_{b,L}^{\rm MBC}$, with bandwidth $b$ and kernel $L$.  
As for the local linear estimator above, we define the corresponding ISE for the MBC estimator as in (\ref{eq:Q_0}) and denote it as $\Delta_{L}^{\rm MBC}(b)$. Its minimizer is the ISE-optimal bandwidth for the MBC estimator with kernel $L$, which we denote as $\widehat{b}_{{\rm ISE},L}^{\rm MBC}$. 

We consider the decomposition
$\widehat{\alpha}_{b,L}^{\rm MBC}(t)-\alpha(t)=B^{\rm MBC}_{b,L}(t)+V^{\rm MBC}_{b,L}(t)$, where $B^{\rm MBC}_{b,L}(t)$ is a stable term converging in probability to zero, and $V^{\rm MBC}_{b,L}(t)$ is a variable term  converging to a Normal distribution. These two terms are defined as follows:
\begin{equation*}\label{var.M}
V^{\rm MBC}_{b,L}(t)=\int f_{t,b}^{\rm MBC}(s)dM(s)
\end{equation*}
where
\begin{eqnarray*}\label{f_MBC}
f_{t,b}^{\rm MBC}(s)&=&\bar{L}^{\rm MBC}_{t,b}(t-s) \frac{\widehat{\alpha}_{b,L}^{\rm LL}(t)}{\widehat{\alpha}_{b,L}^{\rm LL}(s)}+
\bar{L}_{t,b}(t-s)- \\
\nonumber \quad \quad \quad &&\int_0^T\bar{L}^{\rm MBC}_{t,b}(t-u)\frac{\widehat{\alpha}_{b,L}^{\rm LL}(t)}{\widehat{\alpha}_{b,L}^{\rm LL}(u)}\bar{L}_{u,b}(u-s)Y(u)du
\end{eqnarray*}
with $\bar{L}^{\rm MBC}_{t,b}(t-s)$ defined as in ($\ref{K.M}$) for the kernel $L$, 
and
\begin{eqnarray*}\label{bias.M}
B^{\rm MBC}_{b,L}(t)&=&{B}_{b,L}^{\rm LL}(t)+\int \bar{L}^{\rm MBC}_{t,b}(t-s)\widehat{\alpha}_{b,L}^{\rm LL}(t) (\widehat{\alpha}_{b,L}^{\rm LL}(s))^{-1} {B}_{b,L}^{\rm LL}(s) Y(s) ds\\
\nonumber && \quad =\int \bar{L}^{\rm MBC}_{t,b}(t-s)\widehat{\alpha}_{b,L}^{\rm LL}(t)\left(\beta_{b,L}(t)-\beta_{b,L}(s)\right)Y(s)ds ,
\end{eqnarray*}
with $\beta_{b,L}(s)=\{\widehat{\alpha}_{b,L}^{\rm LL}(s)\}^{-1} {B}_{b,L}^{\rm LL}(s)$,  where $B_{b,L}^{\rm LL}$ and $V_{b,L}^{\rm LL}$ are the stable and variable terms for the local linear estimator given in (\ref{B.LL}) and (\ref{V.LL}), respectively.

Using the above decomposition and standard martingale theory along with the approach of \cite{Mammen:Nielsen:07} we can expand the ISE for the MBC estimator. The derivations are close to the local linear case. In Lemma 7 in the supplementary material we show that, under some regularity assumptions,  $\Delta_{L}^{\rm MBC}(b)$ is asymptotically equivalent to 
\[
M_{{L}}^{\rm MBC}(b)=b^{8}\frac{\mu_2^4(\bar{L}^{*})}{16} \int \left\{h(t)\right\}^2\gamma(t)w(t)dt+(nb)^{-1}R\left(\Gamma_{\bar{L}^*}\right)\int \alpha(t)w(t)dt,
\]
with $h(t)=\alpha(t)\left(\alpha''(t)/\alpha(t)\right)''$. From this approximation a deterministic optimal bandwidth for the MBC estimator with kernel $L$ is defined as

\begin{equation}\label{bMISE.MBC}
b^{\rm MBC}_{\rm MISE,L}=C^{\rm MBC}_{0,L} n^{-1/9}; \hskip 2cm C^{\rm MBC}_{0,L}=\left[\frac{R\left(\Gamma_{\bar{L}^*}\right) \int \alpha(t)w(t)dt}{ \frac{\mu_2^4(\bar {L}^*)}{2} \int \left\{h(t)\right\}^2\gamma(t)w(t)dt}\right]^{1/9},
\end{equation}
where $\Gamma_{\bar{L}^*}(u)=2\bar{L}^*(u)-\bar{L}^*(u)\ast\bar{L}^*(u)$ is the kernel obtained by twicing the equivalent kernel, $\bar{L}^*$, given in (\ref{KLequiv}). 

The following theorem states the asymptotic normality of the three bandwidth estimates as well as the infeasible MISE-optimal bandwidth, for the MBC estimator defined with a kernel $K$. The proof is provided in the supplementary material.

\begin{theorem}\label{Th:multi}
Under assumptions A1, A2' and A3', the bandwidth selectors $\widehat{b}_{{\rm BO},K}^{\rm MBC}$,  $\widehat{b}_{{\rm DO},K}^{\rm MBC}$, $\widehat{b}_{{\rm CV},K}^{\rm MBC}$, and $\widehat{b}_{{\rm MISE},K}^{\rm MBC}$,  for the MBC estimator with kernel $K$ satisfy

\begin{eqnarray*}
n^{3/18}\left(\widehat{b}_{{\rm BO},K}^{\rm MBC}-\widehat{b}_{{\rm ISE},K}^{\rm MBC}\right) &&\longrightarrow N\left(0,S_{2}^{\rm MBC}+S_{1}^{\rm MBC} \Psi_{{\rm BO},K}^{\rm MBC}\right) \\
n^{3/18}\left(\widehat{b}_{{\rm DO},K}^{\rm MBC}-\widehat{b}_{{\rm ISE},K}^{\rm MBC}\right) &&\longrightarrow N\left(0,S_{2}^{\rm MBC}+S_{1}^{\rm MBC} \Psi_{{\rm DO},K}^{\rm MBC}\right) \\
n^{3/18}\left(\widehat{b}_{{\rm CV},K}^{\rm MBC}-\widehat{b}_{{\rm ISE},K}^{\rm MBC}\right) &&\longrightarrow N\left(0,S_{2}^{\rm MBC}+S_{1}^{\rm MBC} \Psi_{{\rm CV},K}^{\rm MBC}\right) \\
n^{3/18}\left(\widehat{b}_{{\rm MISE},K}^{\rm MBC}-\widehat{b}_{{\rm ISE},K}^{\rm MBC}\right) &&\longrightarrow N\left(0,S_{2}^{\rm MBC}+S_{1}^{\rm MBC} \Psi_{{\rm MISE},K}^{\rm MBC}\right)
\end{eqnarray*}
where
\begin{eqnarray*}
S_{1}^{\rm MBC}&=& \frac{2^{1/3}}{9^2}\frac{R\left(\Gamma_K\right)^{-15/18}   \left(\int \alpha^2 (t) w^2(t) \ dt \right)}
{\mu_2(K))^{12/9}\left(\int  h(t)^2 \gamma(t) w(t) \ dt \right) ^{\frac{3}{9}}  \left(\int \alpha (t) w(t) \ dt \right) ^{-15/9} } , \\
&& \\
&& \\
S_{2}^{\rm MBC}&=&\frac{2^{30/9}}{9^2}
\frac{ R\left(\Gamma_K\right)^{-6/9} \left(\int h (t)^2 \gamma(t) w^2(t) \alpha(t) \ dt \right)}
{ \left(\mu_2(K)\right)^{12/9} \left(\int \alpha (t) w(t) \ dt \right) ^{6/9} \left(\int h(t)^2 \gamma(t) w(t) \ dt \right) ^{12/9}},
\end{eqnarray*}
with $h(t)= \left\{\frac{\alpha''(t)}{\alpha(t)}\right\}''\alpha(t)$, and
\begin{eqnarray*}
 \Psi_{{\rm BO},K}^{\rm MBC}&=&  \Psi_{{\rm DO},K}^{\rm MBC} = \int {\left \{\frac{R\left(\Gamma_K\right)}{R\left(\Gamma_{\bar K_{\rm L}^*}\right)}\left(H_{\Gamma_{\bar K_{\rm L}^*}} - G_{\Gamma_{\bar K_{\rm L}^*}}\right)(\rho^{\rm MBC} u)-H_{\Gamma_K}(u)\right\}^2}   \ du , \\
\Psi_{{\rm CV},K}^{\rm MBC}&=&  \int \left\{G_{\Gamma_{K}}( u)\right\}^2  \ du  ,\\
\Psi_{{\rm MISE},K}^{\rm MBC}&=&  \int \left\{H_{\Gamma_{K}}( u)\right\}^2  \ du  .
\end{eqnarray*}
where $G_{L}$ and $H_{L}$ are defined as in Theorem \ref{Th:local}, taking  $L=\Gamma_K$ and $L=\Gamma_{\bar K_{\rm L}^*}$. 
\end{theorem}

\begin{Remark}
The result above shows that  all bandwidth selectors have similar asymptotics with the only difference of the factor $\Psi_{\cdot,K}^{\rm MBC}$. A similar conclusion was derived for the local linear estimator.  The three last columns of Table \ref{tab:factors} show the value of this factor for three common choices of $K$.  
\end{Remark}

\section{Two case studies}
\label{sec:data}

In this section the methods proposed in this paper are illustrated with two real data applications. The first application  is on fitting hazard mortality curves for old-age population, and the second one is a non-standard forecasting problem that arises in  non-life insurance.

\subsection{Old-age mortality} \label{sec:mort}

We consider mortality data of women in Iceland in the calendar year 2006, with ages from 40 to 110. The same data were considered by \cite{Gamiz:etal:16} and are available in the \texttt{DOvalidation} R-package (\cite{Gamiz:etal:17}). The data were obtained from the Human Mortality Database and consist of aggregated yearly occurrences and exposures. 
 \cite{Gamiz:etal:16} showed that estimating the hazard from these data is challenge at the oldest ages. The lack of exposure at the right end and the few observed deaths induce a marked boundary effect precisely in the area of interest, the old ages.  
For these data we have calculated the two hazard estimators described in this paper, local linear and MBC, using three bandwidth selectors: cross-validation, double one-sided cross-validation and the new best one-sided cross-validation. The cross-validation scores involved in these methods have been  defined using a weighting function such that $w(s) Y_i(s) \equiv 1$, so all points in the time interval where the hazard function is estimated are evaluated with the same weight. This is different from \cite{Gamiz:etal:16} where the weighting function was chosen so only areas where the exposure is significant contribute to the criteria. Notice that this makes an important difference in this data set where the end of the time interval comprises almost no exposure. 

Before looking at the resulting hazard estimates we shall look at the cross-validation scores to be minimized for each bandwidth selection method. Figure \ref{Fig:scoresMBC} shows the cross-validated scores for each method considering the MBC estimator. The local linear case looks quite similar and can be found in the supplementary material. From these plots we can see that the left one-sided score is not well behaved for both hazard estimators. Therefore the average DO-validated bandwidth becomes unreliable, even though the obtained values seem to be sensible ($\widehat b_{{\rm DO}}=27.3$ for the local linear estimator and $\widehat b_{{\rm DO}}=40$ for the MBC estimator). On the other hand the best one-sided cross-validation method shows a clear minimum in both cases and, as expected, it moves close to the one-sided cross-validated bandwidth that is working fine (the right side in this case).  Best one-sided cross-validation in this case has been calculated using the exposure process, that is, for each time $t$ we use the function $\xi_b^E(t)$ given in (\ref{xi.E}). However the results are quite similar using the occurrence process instead.  Figure \ref{Fig:hazard} shows the resulting hazard estimates from each method and type of hazard estimate. Note from these plots that the MBC hazard estimator is more robust to the bandwidth choice than the local linear estimator. Also the new best one-sided cross-validation method seems to provide a reasonable estimate for old-age mortality in both cases.



\begin{figure}[h]
\centering
\makebox{ \includegraphics[width=10cm]{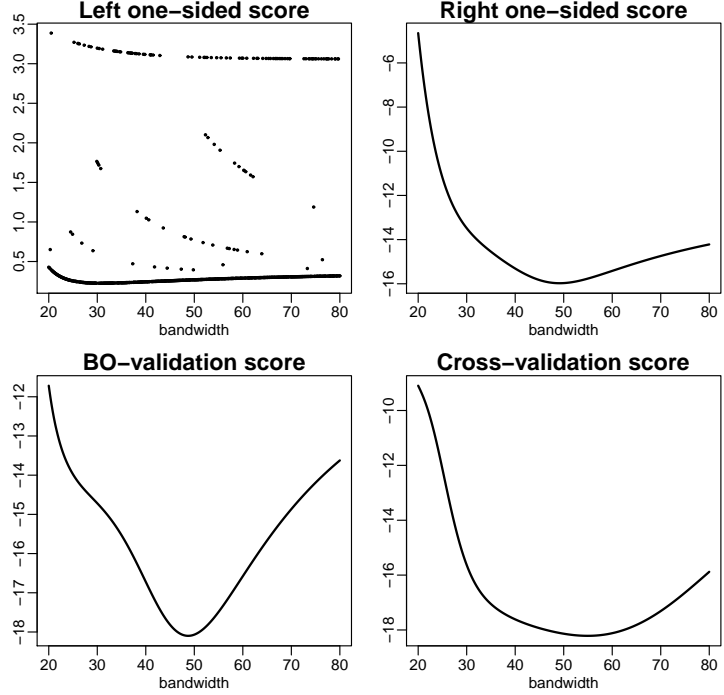}}
\caption{\label{Fig:scoresMBC} Mortality data: bandwidth selection scores with MBC hazard estimator.  }
\end{figure}

\begin{figure}[h]
\centering
\makebox{ \includegraphics[width=10cm]{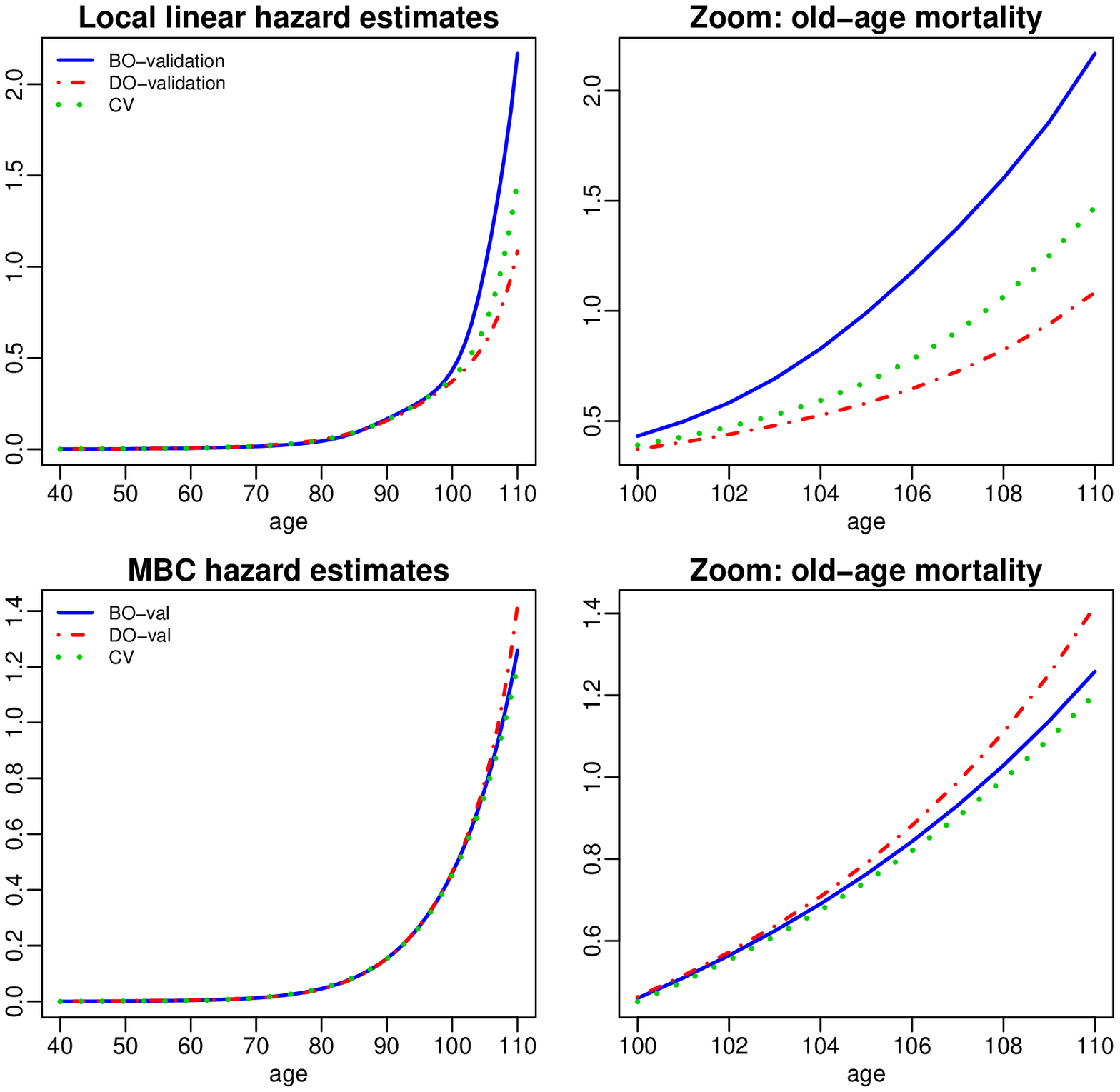}}
\caption{\label{Fig:hazard} Comparison of hazard estimates from female mortality data in Iceland.  }
\end{figure}

\subsection{Outstanding liabilities forecasting in non-life insurance}

The second  application arises in non-life insurance and the goal is to forecast the number of future claims from contracts underwritten in the past, which have not yet  been reported. Typically actuaries are responsible of getting these forecasts, which represent perhaps the most important number in the accounts of the company (see \cite{Martinez:etal:13} for a detailed background of this problem). Here we analyse a data set of  reported and outstanding claims from a motor business in UK. The same data set was previously considered by \cite{Martinez:etal:13} and consist of  $n=1558$ large claims reported between January 1990 and March 2012. From a statistical perspective the data could be described as a sample $\{(X_1,Z_1),\ldots, (X_n,Z_n)\}$, where $X_i$ denotes the underwriting date of the $i$th claim,  and $Z_i$ the corresponding reporting delay, this is, the time between the underwriting date and the  reporting date of the claim. The sample is right truncated since it can be observed only those claims for which the underwriting time plus the reporting delay is not greater than the calendar time of data collection. Hence data exist on a triangle with $X_i + Z_i \leq 31 \text{ March } 2012$, and  $X_i + Z_i$ represents the calendar time. The aim is to forecast the mass of the unobserved, future triangle, where $X_i + Z_i > 31 \text{ March } 2012$, which corresponds to the number of claims underwritten in the past which have not been reported yet. The problem is formulated assuming that the maximum reporting delay  is 267 months, in the actuarial literature this assumption is described as the triangle is fully run off. 
Another challenge of the data set for this problem is that the data are only available in an aggregated way. This is a common feature of this kind of data in the reserving departments of the insurance companies. This means that the available observations are counts living in a triangle of dimension $267\times267$. Specifically for our data set the triangle has entries $\mathcal N_{x,z}=\sum^n_{i=1}  I\big(X_i=x,\  Z_i =z\big), \quad (x, z)\in \{1,\dots,267\}^2$, describing the number of claims underwritten in the $x$th month and reported in the $z$th month. 

\cite{Martinez:etal:13} showed that a multiplicative structured density model, $f(x,z)=f_1(x)f_2(z)$,   can be used to  forecast the claims where  the  components $f_1$ and $f_2$ are  the underwriting time density and the reporting time density, respectively. The assumption of a multiplicative density  means that  the reporting delay does not  depend on the underwriting date.  Using the counting process formulation considered in this paper, \cite{Hiabu:etal:16} solved the forecasting problem estimating the two density components using a time-reversal approach. Data are transformed to the time reversed scale so the right-truncation problem is replaced by the more tractable left-truncation (see \cite{Hiabu:etal:16}, for more details). 
Using the same time-reversal approach, we now use the hazard estimation methods presented in the previous sections to estimate the backward hazard functions corresponding to the two components, underwriting ($\alpha_1$) and reporting delay ($\alpha_2$).  From these hazard estimates the density component estimates can be derived multiplying by respective estimators of the survival functions. 

From the above description we solve the forecasting problem considering both local linear and MBC hazard estimators. For each hazard component, the bandwidth parameters for these estimators have been estimated using cross-validation, double one-sided cross-validation and best one-sided cross-validation. In the three cases we use weighting functions for the involved cross-validation scores that are appropriate for the forecasting problem. Specifically, following the discussion in \cite{Hiabu:etal:16},  to estimate $\alpha_1$ we  consider  weights $w_1(t)=\widehat{S}_1^2(t)\left(1-\widehat{S}_2(t)\right)^2/Y_1(t)$, where $\widehat{S}_1$ and $\widehat{S}_2$ are estimators of the survival functions of each component (underwriting time and the reporting time delay)  on the reversed time scale; and $Y_1(t)$ is the risk process for the first component. In a similar way we define the weights to estimate $\alpha_2$. As in the mortality study best one-sided cross-validation has been calculated using the exposure process. 

Figure \ref{Fig:cashflow} shows the forecasts of the number of claims reported in the future calendar months. Table \ref{Tab:cashflow} shows these forecasts aggregated in years. The forecasts are given for each hazard estimator and bandwidth estimate. We have also included the forecasts derived from the  Chain Ladder method, which involves histogram type estimators of the underwriting and reporting density components. The Chain Ladder method is the classical approach used in the insurance companies (see \cite{Martinez:etal:13} for more details about this approach). The plot of the forecasts shows that the classical insurance method Chain Ladder is overestimating the liabilities, while the kernel hazard methods provide lower forecasts. Previous empirical analyses with these data described in \cite{Martinez:etal:13} agree with this result and recommend multiplicatively bias corrected local linear estimators for this kind of data. 
Looking at the results from the kernel estimators we can see that double one-sided cross-validation and best one-sided cross-validation provide similar forecasts when the local linear estimator is considered, but the results are quite different for the MBC estimator. The predicted total number of claims using DO-validated bandwidth is about 299 compared to 313 using the BO-validated bandwidth. Our concern is that  double one-sided cross-validation might not be behaving properly in this situation. A close inspection to the cross-validation scores to be minimized in order to derive these bandwidth estimates reveals what is happening. 
Figures \ref{Fig:scoresMBC_claim_comp1} and \ref{Fig:scoresMBC_claim_comp2} show these cross-validation scores when the MBC hazard estimator is considered for both underwriting  and reporting delay components. 
From these plots we can see that the right one-sided score completely  breaks down for the underwriting time component, exhibiting several local minima. For the reporting delay component the score function continues decreasing as the value of the bandwidth increases, so it  reaches the minimum at the upper limit of the search interval of bandwidths. The left one-sided score behaves more reasonably for the underwriting component but again breaks down for the reporting delay component. This means that one shouldn't trust the double one-sided cross-validation bandwidth derived from these two one-sided criteria, even though the derived estimates in this case turned to be reasonable values, $\widehat h_{{\rm DO}}=55.8$ for the underwriting time, and $\widehat h_{{\rm DO}}=31.6$ for the delay. On the contrary, the new best one-sided cross-validation method provides bandwidth estimates of $\widehat h_{{\rm BO}}=43.4$ for the underwriting time and $\widehat h_{{\rm BO}}=11.8$ for the delay, exhibiting well-behaved minimization scores as shown in Figure \ref{Fig:scoresMBC_claim_comp1}. 
Regarding  to the  cross-validation method it exhibits a rather flat score in the underwriting component leading to the large bandwidth estimate of $\widehat h_{{\rm CV}}=63.5$, and a value of $\widehat h_{{\rm CV}}=11.6$ for the delay that is close to the best one-sided cross-validated bandwidth. The impact of the cross-validated bandwidths on the forecasts is not significant though, about 309 predicted claims compared to the 313 from best one-sided cross-validation. We have performed the same inspection with the local linear estimators. These plots can be seen in the supplementary material. The picture is again quite similar showing a poor performance of double one-sided cross-validation, however the impact on the forecasts in this case is not substantial. The total number predicted from cross-validation  is about 293, compared to 295 from double one-sided cross-validation and 298 for best one-sided cross-validation. 

%
%

\begin{table}[h]
\caption{\label{Tab:cashflow} Forecasts of the number of claims to be reported in the future calendar years.}
	\centering
{
\begin{tabular}{rrrrrrrr}
		\\\hline\noalign{\smallskip}
 Year & CLM & LL-CV & LL-DO & LL-BO & MBC-CV & MBC-DO & MBC-BO \\ 
  \hline
2012 & 99.95 & 76.85 & 77.98 & 77.95 & 80.55 & 81.75 & 81.76 \\ 
  2013 & 97.23 & 75.06 & 75.52 & 76.86 & 81.18 & 75.82 & 81.68 \\ 
  2014 & 74.32 & 58.75 & 59.05 & 60.04 & 62.23 & 58.89 & 62.88 \\ 
  2015 & 49.18 & 38.88 & 39.06 & 39.44 & 40.31 & 38.81 & 41.20 \\ 
  2016 & 24.52 & 19.42 & 19.50 & 19.66 & 20.01 & 19.34 & 20.44 \\ 
  2017 & 11.61 & 9.35 & 9.39 & 9.44 & 9.60 & 9.45 & 9.76 \\ 
  2018 & 6.21 & 5.07 & 5.06 & 5.09 & 5.15 & 4.99 & 5.27 \\ 
  2019 & 3.24 & 2.54 & 2.53 & 2.52 & 2.52 & 2.53 & 2.61 \\ 
  2020 & 1.36 & 1.25 & 1.23 & 1.23 & 1.23 & 1.18 & 1.22 \\ 
  2021 & 0.99 & 1.03 & 1.02 & 1.02 & 0.99 & 0.95 & 0.95 \\ 
  2022 & 1.11 & 0.85 & 0.84 & 0.85 & 0.87 & 0.83 & 0.88 \\ 
  2023 & 1.06 & 0.71 & 0.71 & 0.73 & 0.81 & 0.80 & 0.85 \\ 
  2024 & 1.20 & 0.81 & 0.81 & 0.84 & 0.93 & 0.90 & 0.94 \\ 
  2025 & 1.14 & 0.91 & 0.92 & 0.95 & 0.97 & 0.93 & 0.94 \\ 
  $>2025$ & 1.94 & 1.59 & 1.59 & 1.61 & 1.51 & 1.48 & 1.55 \\
     \hline\noalign{\smallskip}
     Total &375.07 &293.07& 295.20& 298.23& 308.86& 298.69& 312.92\\ \hline
\end{tabular}
}
	\end{table}
\spacingset{1.5}

%


\begin{figure}[h]
\centering
\makebox{ \includegraphics[width=10cm]{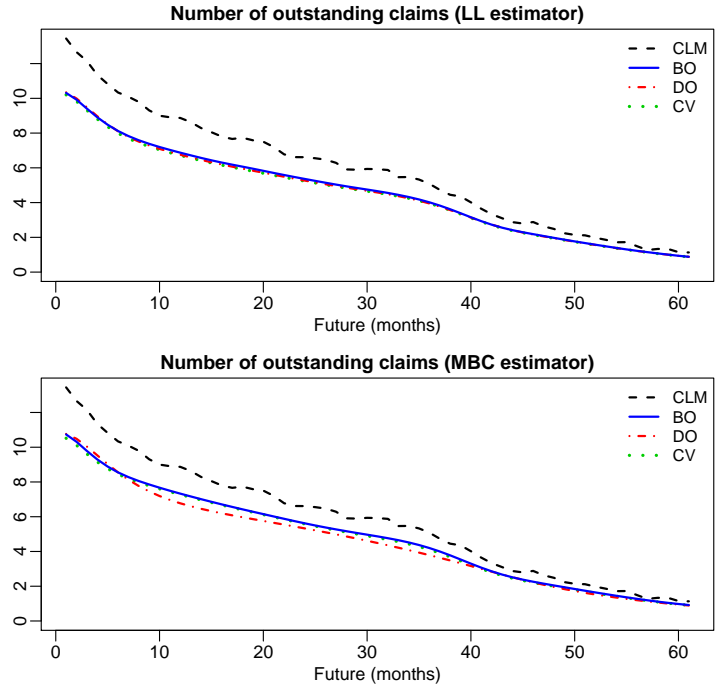}}
\caption{\label{Fig:cashflow} Number of outstanding claims forecast using local linear and MBC estimators.  }
\end{figure}

\begin{figure}[h]
\centering
\makebox{ \includegraphics[width=10cm]{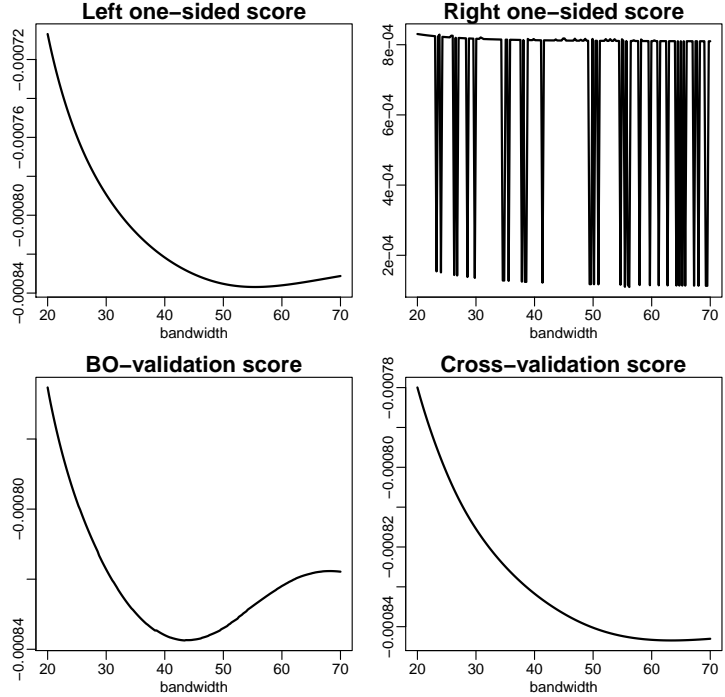}} 
\caption{\label{Fig:scoresMBC_claim_comp1} Underwriting component: bandwidth selection scores with MBC estimator.  }
\end{figure}

\begin{figure}[h]
\centering
\makebox{ \includegraphics[width=10cm]{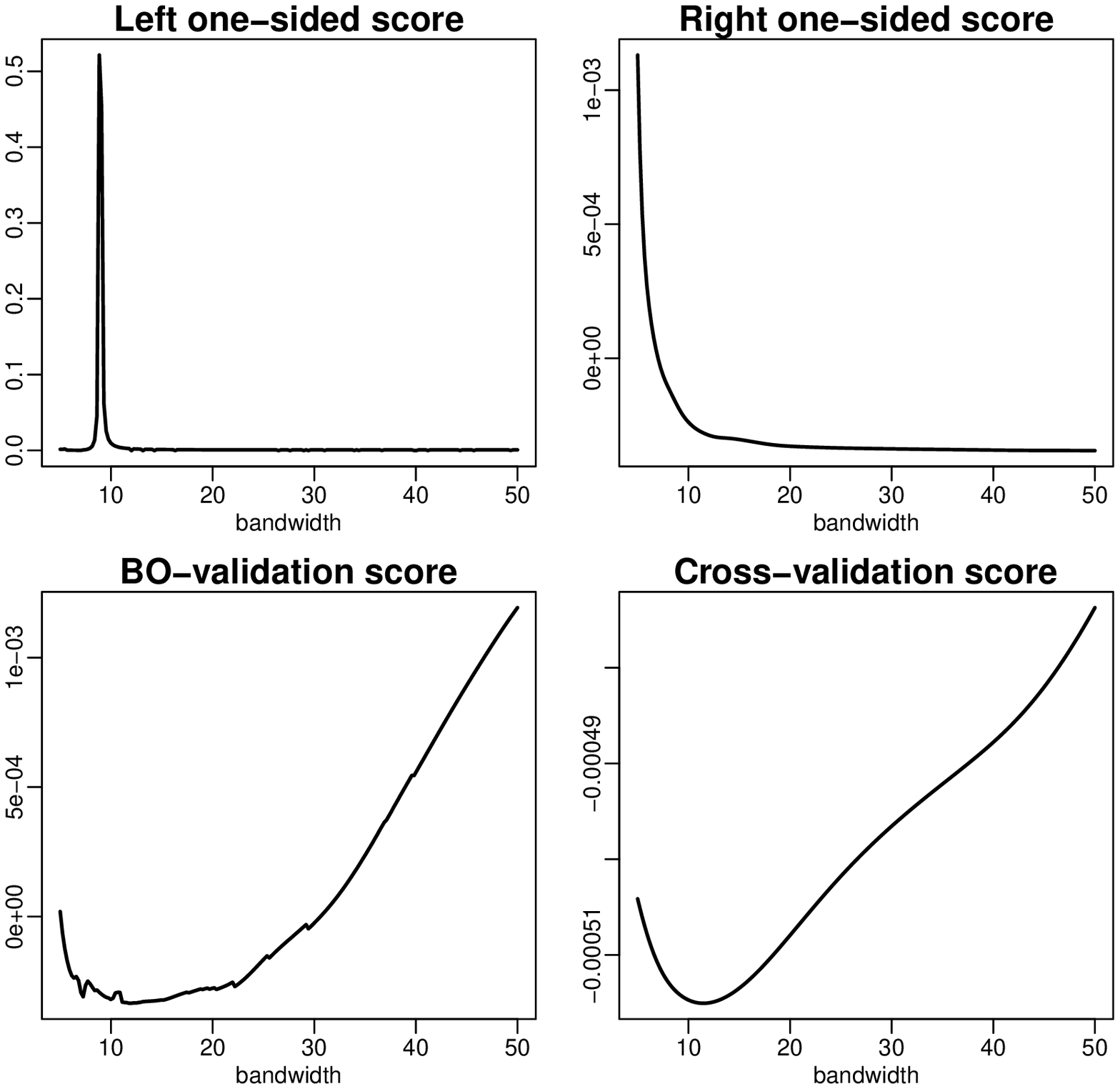}}
\caption{\label{Fig:scoresMBC_claim_comp2} Reporting delay component: bandwidth selection scores with MBC estimator.  }
\end{figure}

\section{Finite sample performance}
\label{sec:sim}

In this section we evaluate the finite sample performance of the new best one-sided cross-validation method for the MBC and the local linear estimators. 
We have considered the same five hazard models described in \cite{Gamiz:etal:16} (see also supplementary material).  
The first four models consist of mixtures of Beta densities. Model 5 shows an exponential decay common in hazard mortality rates as those described in the first case study of Section \ref{sec:data}. From each model we have simulated samples with three different sample sizes and two sampling schemes, right censoring with and without left truncation.  For models 1 to 4 we have considered sample sizes $n=100,1000, 10000$, and for model 5, $n=50000, 75000,100000$. The number of Monte Carlo replications for each case has been always $500$.    We use the same mechanism to simulate data as in  \cite{Gamiz:etal:16}. It generates data in aggregated form (number of occurrences and exposure) for an equally-spaced grid of size $R$ defined on the time interval, and always produces right censored samples. 
 For models 1 to 4 the time interval is $(0, 1)$ and we have defined the grid length with $\delta_R =1/(R +1)$. For model 5 time lies in the interval $(40,110)$ and we have defined the grid length with  $\delta_R =70/(R +1)$.   The  grid size has been chosen equal to $R=500$ in both cases. We shall denote the grid points by $t_r$ ($r =1, \ldots ,R$). In the case of samples without left truncation, for a sample of $n$ individuals, the number of occurrences at time $t_r$, denoted as $O_r$, have been 
generated from the binomial distribution $Bi\left\{Y_r ,\alpha(t_r)\delta_R\right\}$, for $r=1,\ldots ,R$. Here $Y_r$ denotes the size of the risk set at the beginning of the $r$-th interval of the grid. The total number of simulated
occurrences does not sum to $n$. Some of the simulated individuals are finally right censored, because they are still at risk at the end of the interval. Therefore our simulated sample are right censored and the  censoring rates are around 20--30\% for all models. When adding left truncation, independent truncation times  are generated from the uniform distribution.

From the simulated aggregated data we have calculated the local linear and the MBC hazard estimators using the sextic kernel: $K(x)=3003/2048 (1-x^2)^6 I(-1<x<1)$, as in the two data analyses above. For each hazard estimator we have  compared the best one-sided cross-validated bandwidth with cross-validation and  double one-sided cross-validation. 
The performance of the bandwidth estimates have been analysed with respect to the (Monte Carlo approximated) MISE of the resulting kernel hazard estimator. We shall refer to this performance measure as empirical MISE, denoted as $m_1(\widehat b)$,  for each bandwidth estimate $\widehat b$.  As benchmarks in our analysis we have considered two infeasible optimal bandwidths: the ISE-optimal bandwidth minimizing the ISE criterion,  $\widehat b_{{\rm ISE}}$, and the MISE-optimal bandwidth minimizing the empirical MISE.  To compute all bandwidth estimates we have considered grids of 100 equally spaced bandwidth values chosen around the ISE-optimal bandwidth, for each model and sample size. All criteria (ISE, MISE and the cross-validation scores) are defined using a weighting function such that $w(s) Y_i(s) \equiv 1$, so all points in the time interval where the hazard function is estimated are evaluated with the same weight. As we pointed out in our first case study this is different from \cite{Gamiz:etal:16}, and it makes an important difference in models such as Model 5 where the end of the time interval comprises almost no exposure. 

Table \ref{tab:rel.trun} summarizes the simulation results in the case of samples with right censoring and left truncation. In this table bandwidth estimates are compared according to measure $m_1$. For convenience we report a relative measure to indicate when best one-sided cross-validation  outperforms cross-validation. The relative measure is defined as:
\begin{equation*}\label{Rel_err}
Rerr(BO)=\left [m_1(\widehat b_{{\rm CV}})-m_1(\widehat b_{{\rm ISE}})\right ] \big/ \left [m_1(\widehat b_{{\rm BO}})-m_1(\widehat b_{{\rm ISE}})\right ].
\end{equation*}
With this definition values of $Rerr(BO)$ above 1 indicate that best one-sided cross-validation outperforms cross-validation. An analogous  relative measure, $Rerr(DO)$, has been defined for  double one-sided cross-validation. Notice that  $Rerr(BO)$ greater than  $Rerr(DO)$  indicates that best one-sided cross-validation outperforms double one-sided cross-validation.
An overall view of the numbers in the table confirms that best one-sided cross-validation for the multiplicative hazard estimator always outperforms cross-validation, exhibiting  $Rerr(BO)$ values above 1, and double one-sided cross-validation for all models except for few cases where double one-sided cross-validation provides slightly lower empirical MISE values. The results for the local linear estimator show that double one-sided cross-validation and best one-sided cross-validation behave quite similarly, both outperforming in general cross-validation. 
The case of samples without left truncation is shown in Table \ref{tab:rel.cen}. It brings similar conclusions though in this case best one-sided cross-validation is  beaten by double one-sided cross-validation for Model 5. This case deserves a deeper analysis and it is shown in  Table $\ref{tab:mod5}$. In this table we have included the left and right one-sided cross-validated bandwidths (denoted by ``OSCV-l'' and ``OSCV-r'', respectively) from which double one-sided cross-validation is derived. From these results we can clearly see  that the left one-sided bandwidth  completely breaks down, for all sample sizes and both hazard estimators, while the right side behaves well. The average that double one-sided cross-validation performs just hides the problem of the left side. Recall that we pointed  the same issue in the two case studies  described previously. We can see that double one-sided cross-validation was just ``lucky''. On the other hand best one-sided cross-validation is behaving as the best of the two sides, as we would expect.   A similar picture can be seen when analysing the behaviour of double one-sided cross-validation for Model 4 in the case of truncated samples. The full simulation results are provided in the supplementary material. 

\begin{table}[h]
\caption{\label{tab:rel.trun} Simulation results for datasets with right censoring and  left truncation. The relative measure $Rerr$ defined in (\ref{Rel_err}) is shown for BO-validation and DO-validation with local linear and MBC hazards.}
\centering
\begin{tabular}{rrrrrrr}
		\\\hline\noalign{\smallskip}
Model & $n$ & LL-DO & LL-BO & MBC-DO & MBC-BO\\ 
\hline 
1& $100$  &  1.55 & 1.25 & 1.47 & 1.79  \\
&1000  		&	 2.32 & 2.00 & 0.97 & 2.88 \\
&10000 		&  1.90 & 1.71 & 1.82 & 3.30 \\     
2& 100    &  2.28 & 2.04 & 0.46 & 2.47  \\
&1000    	&	 2.42 & 1.99 & 0.15 & 3.66 \\
&10000   	&  2.18 & 1.84 & 0.34 & 3.81 \\     
3& 100    &  1.86 & 1.74 & 1.47 & 1.27  \\
&1000    	&	 0.96 & 0.99 & 0.82 & 1.19 \\
&10000   	&  2.20 & 2.07 & 2.12 & 3.50 \\     
4& 100    &  0.08 & 1.12 & 2.13 & 0.92  \\
&1000    	&	 2.51 & 1.91 & 2.30 & 1.08 \\
&10000    &  2.17 & 1.83 & 3.76 & 2.62 \\     
5 & 50000 &  1.62 & 1.70 & 1.77 & 2.09  \\
&75000    &	 2.04 & 2.18 & 1.41 & 2.31 \\
&$10^5$   &  1.68 & 1.73 & 1.07 & 1.90 \\     
\hline\noalign{\smallskip}
\end{tabular}
\end{table}

\begin{table}[h]
\caption{\label{tab:rel.cen} Simulation results for datasets without left truncation. The relative measure $Rerr$ defined in (\ref{Rel_err}) is shown for BO-validation and DO-validation with local linear and MBC hazards.}
\centering
\begin{tabular}{rrrrrrr}
		\\\hline\noalign{\smallskip}
Model & $n$ & LL-DO & LL-BO & MBC-DO & MBC-BO\\ 
\hline
1& $100$  &  2.58 & 2.05 & 0.89 & 2.51  \\
&1000  		&	 2.62 & 2.27 & 1.24 & 4.60 \\
&10000 		&  2.75 & 2.47 & 1.62 & 8.57 \\     
2& 100    &  2.55 & 1.81 & 0.22 & 2.92  \\
&1000    	&	 2.70 & 2.29 & 0.10 & 3.51 \\
&10000   	&  2.63 & 2.40 & 0.26 & 4.71 \\     
3& 100    &  1.50 & 1.40 & 0.99 & 0.70  \\
&1000    	&	 2.72 & 2.33 & 0.74 & 3.40 \\
&10000   	&  1.81 & 2.10 & 0.65 & 3.52 \\     
4& 100    &  2.03 & 1.89 & 2.19 & 1.13  \\
&1000    	&	 2.09 & 2.03 & 1.28 & 0.90 \\
&10000    &  1.24 & 1.28 & 1.03 & 1.65 \\     
5 & 50000 &  0.80 & 6.45 & 5.33 & 1.60  \\
&75000    &	 0.63 & 5.47 & 4.63 & 1.96 \\
&$10^5$   &  0.56 & 4.32 & 4.16 & 2.28 \\     
\hline\noalign{\smallskip}
\end{tabular}
\end{table}

\begin{table}[h]
	\caption{\label{tab:mod5} DO-validation performance in simulations. The empirical MISE (multiplied by $10^6$) is shown for samples generated from Model 5 without left truncation.}
\centering
\begin{tabular}{lrrrrrrrr}
\\
\hline\noalign{\smallskip}
&\multicolumn{1}{c}{$n$}&\multicolumn{1}{c}{ISE} &\multicolumn{1}{c}{MISE} &
\multicolumn{1}{c}{CV}   &  \multicolumn{1}{c}{OSCV-l} &   \multicolumn{1}{c}{OSCV-r} &  \multicolumn{1}{c}{DO}  &  \multicolumn{1}{c}{BO}   \\ 
\hline\noalign{\smallskip}
Local Linear& 50000 & 1.14 & 2.16 & 13.31  &  \textbf{13424.45} &{ 3.04}& {16.29} &{ 3.03} \\
            & 75000 & 0.82 & 1.32 & 8.82   &  \textbf{4835.00} &{1.92} & {10.28}  &{1.92}\\
            & $10^5$& 0.53 & 0.86 & 4.44  &  \textbf{1894.00}&{1.43} & {7.55} &{1.43} \\
\hline\noalign{\smallskip}
MBC         & 50000 &  0.33 & 0.72 & 11.84  &  \textbf{203897.80} & {7.01}&{2.49} & {7.54} \\
					  & 75000 &  0.23 & 0.43 &  5.74  &  \textbf{ 92670.27} & {3.01} &{1.42}  &{3.04}\\
						& $10^5$&  0.15 & 0.27 &  3.51  &  \textbf{16302.04} &{1.65} &{0.96} & {1.63} \\
\hline\noalign{\smallskip}
\end{tabular}
\end{table}

\section{Conclusion}

We have proposed a new bandwidth selection method for local linear hazard estimation and its multiplicatively bias correction. Our proposal is called best one-sided cross-validation and consists of an improvement of  the double one-sided cross-validation of \cite{Gamiz:etal:16}. Best one-sided cross-validation solves the  lack of stability of double one-sided cross-validation in practice via a local information principle. Our empirical studies show that best one-sided cross-validation provides a good strategy for bandwidth selection for both local linear and multiplicative bias corrected hazard estimators. Best one-sided cross-validation inherits the good properties of one-sided cross-validation while avoiding the stability problems that double one-sided cross-validation sometimes faces. Detailed mathematical theory at the level of \cite{Hall:Marron:87} and \cite{Gamiz:etal:16} is included. This type of theory is completely novel for the multiplicatively bias corrected hazard estimators. Theory on best-one-sided cross-validation introduced in this paper is of course also new for the local linear hazard estimator.

\bigskip
\begin{center}
{\large\bf SUPPLEMENTARY MATERIAL}
\end{center}

\begin{description}

\item[SuppPub1:] contains more details on the asymptotics, including the proofs, as well as additional plots and tables for case studies and simulations. (.pdf file)

\end{description}

%
%

%
\end{document}